\documentclass{article}
\usepackage[utf8]{inputenc}
\usepackage{authblk}
\usepackage{setspace}
\usepackage[top=1in,bottom =1 in ,left=0.5 in,right=0.5 in]{geometry}
\usepackage{graphicx}
\graphicspath{ {./figures/} }
\usepackage{subcaption}
\usepackage{amsmath}
\usepackage{xcolor}
\usepackage{nicefrac}
\usepackage{tikz}
\usetikzlibrary{calc}
\usetikzlibrary{arrows.meta,positioning}
\usetikzlibrary{shapes,arrows}
\usepackage{verbatim}
\usepackage{hyperref}
\usepackage{caption}
\usepackage{subcaption}

\def \gaot{(\gamma+ \alpha+ d+ m_{12})} 
\def \gato{(\gamma+ \alpha+ d+ m_{21})} 
\def \siot{(\sigma +d + m_{12})} 
\def \sito{(\sigma +d + m_{21})}
\def \mot{m_{12}}
\def \mto{m_{21}} 

\usepackage{booktabs,dcolumn,caption}
\newcolumntype{d}[1]{D{.}{.}{#1}} 

\hypersetup{
	colorlinks=true,
	linkcolor=blue,
	filecolor=magenta,
	urlcolor=blue,
	citecolor= blue,
	bookmarks=true,
}

\title{Effect of population migration and punctuated lockdown on the spread of
	infectious diseases}

\author[1]{\small Ravi Kiran\footnote {ravieroy123@gmail.com, ravieroy123@iitkgp.ac.in}}
\author[2]{\small Madhumita Roy\footnote{madhumita.roy@cnci.org.in}}
\author[3,*]{\small Syed Abbas\footnote {abbas@iitmandi.ac.in, $^*$Corresponding author}}
\author[1,4]{\small A. Taraphder\footnote{arghya@phy.iitkgp.ernet.in}}
\affil[1]{\footnotesize \textit{Centre for Theoretical Studies,Indian Institute of Technology Kharagpur, Kharagpur 721302, India}}
\affil[2]{\footnotesize \textit{Department of Environmental Carcinogenesis and Toxicology, Chittaranjan National Cancer Institute, 37 S. P. Mukherjee	Road, Kolkata 700026, India}}
\affil[3]{\footnotesize \textit{School of Basic Sciences, Indian Institute of Technology Mandi HP 175005, India} }
\affil[4]{\footnotesize \textit{Department of Physics, Indian Institute of Technology Kharagpur, Kharagpur 721302, India}}

\date{}

\begin{document}

\maketitle

\begin{abstract}
One of the critical measures to control infectious diseases is a lockdown. Once past the lockdown stage in many parts of the world, the crucial question now concerns the effects of relaxing the lockdown and finding the best ways to implement further lockdown(s), if required, to control the spread. With the relaxation of lockdown, people migrate to different cities and enhance the spread of the disease. This work presents the population migration model for n-cities and applies the model for migration between two and three cities. The reproduction number is calculated, and the effect of the migration rate is analyzed. A punctuated lockdown is implemented to simulate a protocol of repeated lockdowns that limits the resurgence of infections. A damped oscillatory behavior is observed with multiple peaks over a period.
\end{abstract}

{\bf Keywords}: SEIRS modelling; Multi-city model; Population migration; Mathematical model.\\
\vskip .5cm

\section{Introduction}

Currently, the world is very much under the threat of an infectious disease. People with compromised immune systems, and especially old individuals, are at the highest risk from this virus. A global pandemic tests our capability to control the situation and protect the population effectively. For the current ongoing pandemic using estimates for seasonality and immunity,  Kissler \textit{et al.} \cite{kissler2020projecting} projected that prolonged social distancing may be necessary till 2022, and SARS-CoV-2 surveillance should be maintained since a resurgence of the virus could be possible as late as 2024. The primary aim of any modeling is to devise strategies that reduce human loss and predict speedy recovery back to normal life.  To that end, protection or prevention from the spread of the virus is of paramount interest.

It becomes imperative to have some idea about its spread and some ways to control the situation. This can help the health and administrative authorities to assess the situation better and be prepared for any eventuality. Since Kermark and Mckendrick in  \cite{kermark1927contributions} built up a system to study epidemiology in 1927, the concept of “compartment modeling” has been widely used until now. Since then several analytical work \cite{zhang2006global, sun2010global, wang2005age} has been carried out on epidemic diffusion focused on compartmental models. Generally, the population is divided into different classes of individuals, which are allowed to mix homogeneously.  In the present work, we study the SEIRS (Susceptible - Exposed - Infectious - Recovered – Susceptible) model for migration between n-cities. We then analyze a particular problem at hand, namely, the diffusion of the population between two and three cities in the backdrop of the COVID attack. Population diffusion has been introduced in \cite{ahmed2012modeling}, and an earlier, more general study using the SIS model was studied by Liu et al. \cite{liu2013modeling}; we use the SEIRS model as it is more suitable to analyze the present COVID situation. We pose a few questions and suggest control measures for the benefit of the population.  No effect of vaccines has been considered, as it is still not available. In our model, a small fraction of the people in a city migrate to another city. In earlier work in SIR, model \cite{chen2014transmission} the susceptible individuals are allowed to migrate in or out of the cities. In the present work, we enable individuals of all types to migrate between the cities.

One of the crucial problems arising from population mixing is the resurgence of infected and exposed individuals. The novelty of the new virus is that while an individual might not be infected, there is a good chance for them to be susceptible or even exposed. When one of the cities has a large influx of individuals, the possibility of an outbreak increases, and in the absence of any vaccine, the first measure is a strict containment implementation.
The best way to reduce the exposed and infected individuals is a significant question with the parameters at hand. It turns out that this depends on (i) how early we take action and (ii) migration parameters and reinforcement of public health diktats. This also provides us with a long-term solution that can be implemented to bring the situation back to some normalcy.

\section{SEIR Model: Theory and Differential Equation}\label{sec: basic_model }

\subsection{\textit{Notation}}

We summarize the list of variables used in this section as follows:

\begin{table}[!h]
	\centering
	\caption{Table of notation.}
	\label{Table_1}
	\begin{tabular}{ p{2cm} p{15cm}}
		\hline
		\multicolumn{2}{c}{\textbf{Variables}} \\
		\hline
		Notation & Description \\
		\hline
		S & Susceptible population \\
		E & Exposed(latent) population  \\
		I & Infected population \\
		R & The population who recovered from the disease  \\
		\hline
	\end{tabular}
	\begin{tabular}{ p{2cm} p{15cm}}
		\hline
		\multicolumn{2}{c}{\textbf{Fixed Parameters}} \\
		\hline
		Notation & Description \\
		\hline
		A & Birth rate of individuals \\
		$m_{ij}$ & Migration parameter describing migration from city j to city i  \\
		$\beta_i$ & Probability of disease transmission in city i \\
		$\nicefrac{1}{\eta}$ & half-saturation constant of infected individuals.   \\
		$\delta$ & Fraction of the recovered individual who loses their immunity and becomes susceptible in unit time.\\
		$\nicefrac{1}{d}$ & Average life expectancy.  \\
		$\nicefrac{1}{\alpha}$ & Average life expectancy of infected individual.  \\
		$\nicefrac{1}{\gamma}$ & Length of infectious period for the infected population.  \\
		$\nicefrac{1}{\sigma}$ & Length of latent period for the exposed population\\
		\hline
	\end{tabular}
\end{table}

\newpage

\subsection{\textit{SEIR: Basic Modelling}}\label{sec:basic_modelling}

There is a long history of mathematical models in epidemiology, and most of them are compartment models, with the population divided into different classes. This type of compartment model for capturing the dynamics of infectious disease was proposed by Kermack, and McKendrick \cite{kermark1927contributions}. In their model (SIR), the population was divided into three states which have been modified with an addition of another compartment, and many versions of this model (SEIR) can be found, for instance, in Hethcote \cite{hethcote2000mathematics}, Diekmann et al. \cite{diekmann2012mathematical}. SEIR model has the following characteristics:

\begin{itemize}
	\item Susceptible class, S(t) is the number of individuals not yet infected with the disease at time t. These individuals are susceptible to the disease.
	
	\item Exposed class, E(t) are the exposed individuals but not yet infectious.
	
	\item Infected class, I(t) denotes the number of individuals who have been infected with the disease and are capable of spreading disease to those who are susceptibles.
	
	\item Recovered class, R(t) are the individuals who recovered from the infection either due to immunization or due to death.
\end{itemize}

Many contact rate forms are used in the incidence term in the epidemic models, the most common being the standard incidence $\frac{\lambda SI}{N}$ which starts with the assumption that the adequate contact rate is a constant $\lambda$. The bilinear incidence $\beta SI = \beta N \frac SN I$ implies that the proper contact rate $\beta N$ is linearly proportional to the total population size N. The drawback of this is that it would be unrealistic for a population too large. Since we want to simulate some degree of social distancing, we use the saturated incidence rate.

In the present paper, we assume that immunity does not last forever, and a fraction of the population loses their immunity and becomes susceptible again. The following flowchart shows the basic SEIRS model with social distancing.

\begin{center}
	\begin{tikzpicture}[node distance=1cm, auto,
		>=Latex,
		every node/.append style={align=center},
		int/.style={draw, minimum size=1cm}]
		\tikzstyle{input} = [coordinate]
		\node [input, name=input] {};
		\node [int, right=of input] (S) {$S$};
		\node [int, right=of S] (E) {$E$};
		\node [int, right=of E] (I) {$I$};
		\node [int, right=of I] (R) {$R$};

		\coordinate[right=of I] (out);
		\coordinate[below=of S] (sdown);
		\coordinate[below=of E] (edown);
		\coordinate[below=of I] (idown);
		\coordinate[below=of R] (rdown);
		\coordinate[above=of S] (sleft);

		\draw [draw,->] (input) -- node {$A$} (S);

		\path[->]
		(S) edge node {$\frac{\beta I S}{1+\eta I}$} (E)
		(S) edge node {$dS$} (sdown)
		(E) edge node {$dE$} (edown)
		(I) edge node {$(\alpha+d)I$} (idown)
		(R) edge node {$dR$} (rdown)
		(E) edge node {$\sigma E$} (I)
		(I) edge node {$\gamma I$} (R)
		(R) edge[out=120, in=60] node {$\delta R$} (S);
		

	\end{tikzpicture}
\end{center}

The ordinary differential equations can then be formulated as

\begin{align}
	\label{eqn: SEIR_Model}
	\frac{dS}{dt} &= A -\frac{\beta SI}{1+ \eta I}+ \delta R -dS \nonumber \\
	\frac{dE}{dt} &=\frac{\beta SI}{1+ \eta I} -\sigma E -d E \\ \nonumber
	\frac{dI}{dt} &= \sigma E -\gamma I -\alpha I-dI \nonumber\\
	\frac{dR}{dt} &=\gamma I -\delta R -dR \nonumber
\end{align}

\noindent Since the total population is not constant, we have Total population at time t as, $N(t)= S(t) + E(t) +I(t)+ R(t)$.

\subsection{\textit{Population migration model with multiple cities}} \label{sec:pop_mig}
For infectious diseases, the study of migration between different populations is quite important. The study of mobility on the spread of an infectious disease can be accomplished in various ways, such as in \cite{arino2015epidemiological} where the SIR model has been implemented to study migration between the larger city and several smaller cities. In this section, we present the generalized migration between n cities.   We use the SEIRS model described in section \ref{sec:basic_modelling}. We consider a model with state variables $S_{i}$, $E_{i}$, $I_{i}$, $R_{i}$ that represent the number of susceptible, infected, exposed and recovered individuals in $ \text{city}_i$, respectively. The suffix i runs from 1 to n and describes the $\text{i}^{\text{th}}$ city. Table I lists the variables and parameters used, with their suffix referring to the city ($i,j=1\cdots n$).

The parameter $\beta_{i}$ is the transmission coefficient in the city i; it has units per unit time per individual. It represents the rate of infection. The parameter $\gamma$ is the per capita rate of recovery, i.e., $\gamma$ is the length of the infectious period for the infected population. The constant A is the rate of inflow of new individuals either by birth or from some external cities not under consideration. The parameter d is the natural death rate; the reciprocal $\nicefrac{1}{d}$ is interpreted as the average life expectancy (e.g., 70 years), which refers to the average normal deaths not related to infectious disease. The parameter $m_{ij}$ is the rate of movement from city j to i. $\delta$ is the fraction of individuals from the recovered class who become susceptible again. Unless stated otherwise, all the parameters are positive.

The total population of the city i is $N_{i} = S_{i} + E_{i} + I_{i} + R_{i}$. The differential equations governing the migration between cities is:

\begin{align}
	\label{eqn: pop_model}
	\frac{dS_{i}}{dt} &= A - \sum\limits_{\substack{j=1 \\ j\neq i}}^n  m_{ji}S_{i} + \sum\limits_{\substack{j=1 \\ j\neq i}}^n m_{ij}S_{j}-\frac{\beta_{i} S_{i}I_{i}}{1+ \eta_{i} I_{i}}+ \delta  R_{i} -dS_{i} \nonumber \\
	\frac{dE_{i}}{dt} &=\frac{\beta_{i} S_{i}I_{i}}{1+ \eta I_{i}} - (\sigma + d) E_{i} -\sum\limits_{\substack{j=1 \\ j\neq i}}^n  m_{ji}E_{i} + \sum\limits_{\substack{j=1 \\ j\neq i}}^n m_{ij}E_{j} \\ \nonumber
	\frac{dI_{i}}{dt} &=  \sigma E_{i} -(\gamma +\alpha +d )I_{i}-\sum\limits_{\substack{j=1 \\ j\neq i}}^n  m_{ji}I_{i} + \sum\limits_{\substack{j=1 \\ j\neq i}}^n m_{ij}I_{j} \nonumber \\
	\frac{dR_{i}}{dt} &=\gamma I_{i} - \delta  R_{i} - dR_{i} -\sum\limits_{\substack{j=1 \\ j\neq i}}^n  m_{ji}R_{i} +\sum\limits_{\substack{j=1 \\ j\neq i}}^n m_{ij}R_{j} \nonumber
\end{align}

\section{Population Migration: application to two cities} \label{sec: two_city}

As an application of the above model we analyse the system(\ref{eqn: pop_model}) for two cities in this section. The notation used in system(\ref{eqn: pop_model}) is already explained in Table(\ref{Table_1}). The parameters used in the simulation for two cities is given in Table(\ref{Tabel_2}).

\begin{table}[h]
	\centering
	\setlength\tabcolsep{0pt} 
	\caption{Parameters used for simulation.}
	\label{Tabel_2}
	\begin{tabular}{ p{6cm} p{2cm} p{3cm} p{4cm}}
		\multicolumn{4}{c}{\textbf{Fixed Parameters}} \\
		\hline
		Description & symbol & value & Reference \\
		\hline
		Per -capita birth rate          & A      &       10/day & - \\
		Normal life expectancy                    & $\nicefrac{1}{d}$      &      69.41 years & \footnotesize{\url{https://data.worldbank.org/}} \\
		Contact rate                          & $\beta$  &      $\nicefrac{0.25}{\text{day}}$ & \cite{tang2020prediction} \\
		Social distancing parameter           & $\eta $  &       0 - 1 & - \\
		Return to susceptible rate            & $\delta$ &      0.02 & - \\
		Infectious period                       & $\nicefrac{1}{\sigma}$ &      5.2 d & \cite{CoronaTracker} \\
		Virus-induced average death rate            & $\alpha$ &      0.02 & - \\
		Recovery period            & $\nicefrac{1}{\gamma}$ &   8 d    & \cite{carcione2020simulation}  \\
		
		\bottomrule
	\end{tabular}
\end{table}

\subsection{\textit{Description of cities}}\label{sec:description_of_cities}
In order to analyze the effect of migration on the epidemic, we try to make some realistic situations for the two cities. For the first case, we consider migration from a well-developed larger city(L) to a lesser developed smaller city (S). The city L has better infrastructure, thus attracting people from city S. They can also implement better preventive measures in social distancing and control the migration rate. The initial conditions are: For city L (city number 1; $N_1, S_1, E_1, I_1, R_1  = 10^8, N_1-E_1-I_1-R_1, 1000,1,0)$ and for city S (city number 2; $N_2, S_2, E_2, I_2, R_2  = 10^7, N_2-E_2-I_2-R_2, 1000,1,0)$. The contact rate $\beta$ is assumed to be the same for both the cities and $\eta$, which is the saturation
the factor that measures the inhibitory effect and simulates the preventive measures is adjusted accordingly to simulate that one city (1st city) has better social distancing than the other (2nd).

\subsection{Reproduction Number ($R_0$)} \label{sec: R_0}
The basic reproduction number is a fundamental concept in epidemiology. It represents the number of secondary infections resulting from a single primary infection in the susceptible population. When finitely many different categories of individuals are involved, the next-generation matrix(NGM) is the natural basis for the definition of reproduction number. In compartmental models, individuals are in a finite number of discrete states that change with time from one state to another. As is the case, the same symbol is used as a label for a state and to denote the corresponding population size. Ideally, we could use the full system of equations consisting of Susceptible, Exposed, Infected, and Recovered compartments for both the cities to calculate the reproduction number $R_0$. Following the work by Diekmann et al. \cite{diekmann2010construction}, we only need to concern with those compartments which contain infections. The susceptible compartment only allows migration of individuals from the susceptible compartment of other cities. As discussed later in the section, only the states `at-infection' are involved in the action of \textbf{K}, and hence in the computation of $R_0$. So for the computation of $R_0$, we only consider the states that apply to infected individuals, which for the case of two cities are $(E_1, E_2, I_1, I_2)$. For notations used see Table \ref{Table_1}
\begin{align}
	\label{eqn: sub_system}
	\frac{dE_{1}}{dt} &=\frac{\beta S_{1}I_{1}}{1+ \eta_{1} I_{1}} -\sigma E_{1} -d E_{1} -m_{21}E_{1} + m_{12}E_{2}  \nonumber \\
	\frac{dI_{1}}{dt} &= \sigma E_{1} -\gamma I_{1} -\alpha I_{1}-dI_{1}-m_{21}I_{1} + m_{12}I_{2}\nonumber  \\  \\
	\frac{dE_{2}}{dt} &=\frac{\beta S_{2}I_{2}}{1+ \eta_{2} I_{2}} -\sigma E_{2} -d E_{2} +m_{21}E_{1} - m_{12}E_{2} \nonumber \\
	\frac{dI_{2}}{dt} &= \sigma E_{2} -\gamma I_{2} -\alpha I_{2}-dI_{2}+m_{21}I_{1} - m_{12}I_{2} \nonumber
\end{align}

We set $\mathbf{x} = (E_1, I_1, E_2, I_2)^\prime$, where the prime denotes transpose. Then we write the linearized infection subsystem in the form
\begin{equation*}
	\mathbf{\dot{x}} = (\mathbf{T} + \mathbf{\Sigma})\mathbf{x}
\end{equation*}

The matrix \textbf{T} is the \textit{transmission} part, describing the production of new infections and \textbf{$\Sigma$} is the \textit{transition} part, describing the changes in the state. At the infection free steady state $E_1 = E_2= I_1= I_2= R_1= R_2 =0$. We assume disease free equilibrium (DFE) exists and is stable in the absence of disease and detailed proofs of such can be found in \cite{van2002reproduction}. Diekmann \textit{et.al} \cite{diekmann2010construction} gave a method to find the Reproduction number($R_0$) by constructing next-generation matrix(NGM) for the complete system. Following the method outlined by Arino et.al for calculation of reproduction number for multiple cities in \cite{arino2003multi, arino2003basic}, we find $R_0$  for multiple cities
.
Regarding the the linearized subsystem (\ref{eqn: sub_system}) we obtain,
$$
\mathbf{T} =
\begin{pmatrix}
	0 & \nicefrac{\beta A}{m_{21}+d} & 0 & 0\\
	0 & 0 & 0 & 0\\
	0 & 0 & 0 & \nicefrac{\beta A}{m_{12}+d} \\
	0 & 0 & 0 & 0
\end{pmatrix}
$$

and,

$$
\mathbf{\Sigma} =
\begin{pmatrix}
	-(\sigma+d+m_{21}) & 0 & m_{12} & 0\\
	\sigma & -(\gamma +\alpha +d+m_{21}) & 0 & m_{12}\\
	m_{21} & 0 & -(\sigma+d+m_{21})  & 0 \\
	0 & m_{21} & \sigma & -(\gamma +\alpha +d+m_{12})
\end{pmatrix}
$$

The determinant of the matrix $\mathbf{\Sigma}$ is:

\begin{equation*}
	\mathbf{D} =  \left[\left(\alpha +\gamma +d+m_{12}\right) \left(\alpha +\gamma +d+m_{21}\right)+m_{12} m_{21} \right] \left[(\sigma +d+ m_{21})(\sigma +d+ m_{12}) - m_{12}m_{21}\right]
\end{equation*}

Hence we have,

\begin{equation*}
	\mathbf{\Sigma^{-1}} = \frac{1}{\mathbf{D}}
	\begin{pmatrix}
		C_{11} &C_{21} & C_{31} & C_{41}\\
		-C_{12} & -C_{22} & -C_{32} & -C_{42}\\
		C_{13} & C_{23} & C_{33}  & C_{43} \\
		-C_{14} & -C_{24} & -C_{34} & -C_{44}
	\end{pmatrix}
\end{equation*}

Where,

\begin{align*}
	C_{11} &= -(\sigma +d + m_{12})\left[(\gamma+\alpha +d + m_{21})(\gamma+ \alpha+ d+ m_{12})- m_{12}m_{21}\right]\\
	C_{12} &= (\gamma+ \alpha+ d+ m_{12})\left[\sigma (\sigma +d + m_{12}) \right]\\
	C_{13} &= -m_{21}\left[(\gamma+\alpha +d + m_{21})(\gamma+ \alpha+ d+ m_{12})- m_{12}m_{21}\right]\\
	C_{14} &= (\sigma m_{21})\left[(\gamma+ \alpha+ d+ m_{21}) (\sigma +d + m_{12}) \right]
\end{align*}

\begin{align*}
	C_{21} &= 0\\
	C_{22} &= -\gaot \left[\siot \sito - \mot \mto \right]\\
	C_{23} &= 0\\
	C_{24} &= -\mto \left[\siot \sito -\mot \mto   \right]
\end{align*}

\begin{align*}
	C_{31} &= \mot \left[\gato \gaot - \mot \mto  \right]\\
	C_{32} &= \sigma \mot \left[ \gaot + \sito  \right]\\
	C_{33} &= -\sito \left[\gaot \gato - \mot \mto   \right]\\
	C_{34} &= \sigma [\gato \sito + \mot \mto]
\end{align*}

\begin{align*}
	C_{41} &= 0\\
	C_{42} &= -\mot[\sito \siot - \mot \mto]\\
	C_{43} &= 0\\
	C_{44} &= - \gato [\sito \siot - \mot \mto]
\end{align*}

Now the NGM with large domain $\mathbf{K_L}$ is four-dimensional and given by $\mathbf{K_L} = - \mathbf{T} \mathbf{\Sigma^{-1}}$.

\begin{equation*}
	\left(
	\begin{array}{cccc}
		C_{12} \text{$\beta $S}_{10} & C_{22} \text{$\beta $S}_{10} & C_{32} \text{$\beta $S}_{10} & C_{42} \text{$\beta $S}_{10} \\
		0 & 0 & 0 & 0 \\
		C_{14} \text{$\beta $S}_{20} & C_{24} \text{$\beta $S}_{20} & C_{34} \text{$\beta $S}_{20} & C_{44} \text{$\beta $S}_{20} \\
		0 & 0 & 0 & 0 \\
	\end{array}
	\right)
\end{equation*}

Where, $S_{10} $ and $S_{20}$ are susceptible population in disease free case for city 1 and city 2 respectively. Now we formally obtain \textbf{K} from $\mathbf{K_L}$ by pre- and post-multiplying $\mathbf{K_L}$ by an auxiliary matrix \textbf{E} that singles out the rows and columns relevant for the reduced set of states. We specify \textbf{E} as consisting of unit column vectors $\mathbf{e}_i$ for all i such that $\text{i}^\text{th}$ row of \textbf{T} is not identically zero. For the above matrix, we have

\begin{equation*}
	E= \left(
	\begin{array}{cc}
		1 & 0 \\
		0 & 0 \\
		0 & 1 \\
		0 & 0 \\
	\end{array}
	\right)
\end{equation*}

To find the NGM, we then perform the matrix multiplication

\begin{equation*}
	\mathbf{K} = \text{E}^\prime \mathbf{K_L} \text{E}
\end{equation*}
Which gives us,
\begin{equation*}
	\mathbf{K} = \frac1D \left(
	\begin{array}{cc}
		\text{$\frac{ C_{12} \beta A}{\mto +d} $} & \text{$\frac{ C_{32} \beta A}{\mto +d} $} \\
		\text{$\frac{ C_{14} \beta A}{\mot +d} $} & \text{$\frac{ C_{34} \beta A}{\mot +d} $} \\
	\end{array}
	\right)
\end{equation*}

Now, we know that for a $ 2 \times 2$ matrix the dominant eigenvalue , and hence $R_0$ is obtained from the its trace and the determinant as

\begin{align*}
	R_0 &= \rho(\mathbf{K})\\
	& = \frac12 \left(\text{Tr}(\mathbf{K}) + \sqrt{\text{Tr}(\mathbf{K})^2 - 4  \ \text{det}(\mathbf{K})}\right)
\end{align*}

\subsubsection{Variation of $R_0$}

\begin{figure}[!h] 
	\centering
	\includegraphics[scale=1]{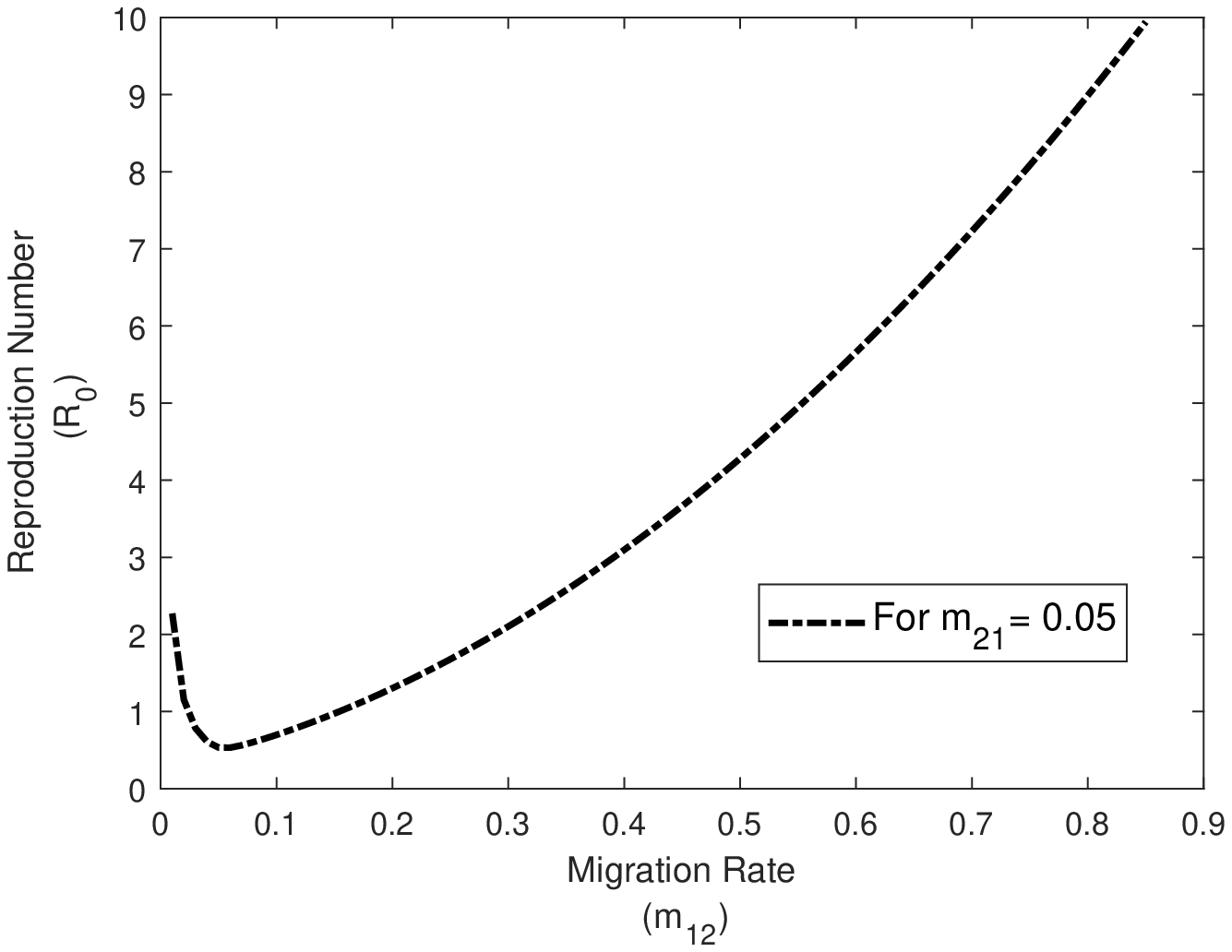}
	\caption{Variation of $R_0$ with $m_{12}$ for $m_{21} = 0.05$.}
	\label{fig:R0_m12}
\end{figure}%

$R_0$ is a vital parameter affected most by the migration between cities; we see the variation of reproduction number with migration parameter.
For parameters mentioned in Table (\ref{Tabel_2}) and using the formula derived for reproduction number for two cities, fig.(\ref{fig:R0_m12}) shows the variation of $R_0$ with migration rate.

The increase of migration rate increases the reproduction number for the whole system for migration from either city. The slight dip in reproduction number in fig.(\ref{fig:R0_m12}) is because we started with a larger migration rate $m_{21}$. This supports the assertion that migration between cities can lead to an increase in the spread of the disease.

\subsection{Numerical Simulations}\label{numerical_sims}
\begin{figure}[h]
	\centering
	\begin{subfigure}{.5\textwidth}
		\centering
		\includegraphics[scale=0.55]{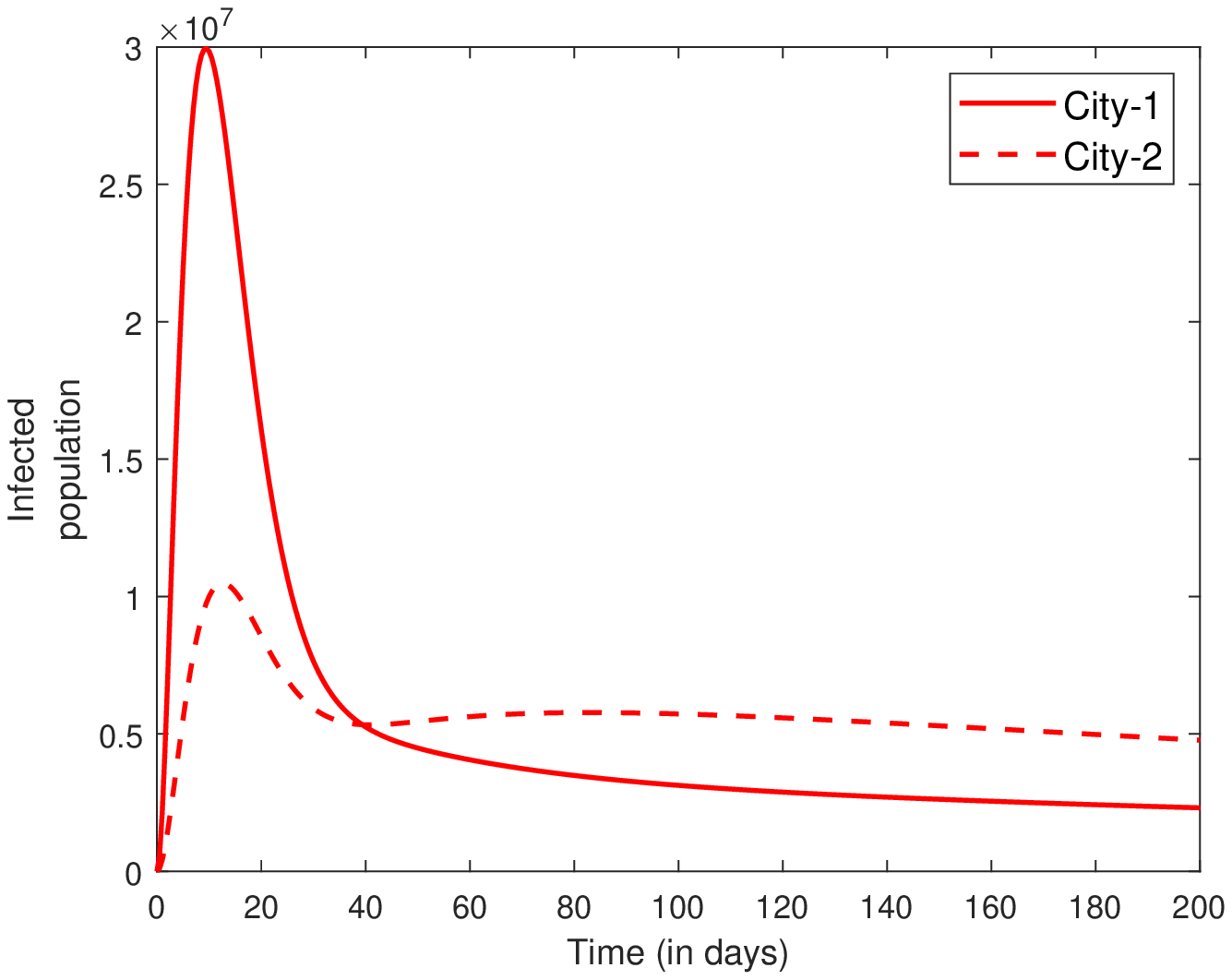}
		\caption{evolution of infected population. The maximum infected population for city 1 and city 2: $(I^{m}_{1}, I^{m}_{2}) = (2.99 \times 10^7 , 1.04 \times 10^7)$}
		\label{fig:inf_mig_1}
	\end{subfigure}%
	\begin{subfigure}{.5\textwidth}
		\centering
		\includegraphics[scale=0.55]{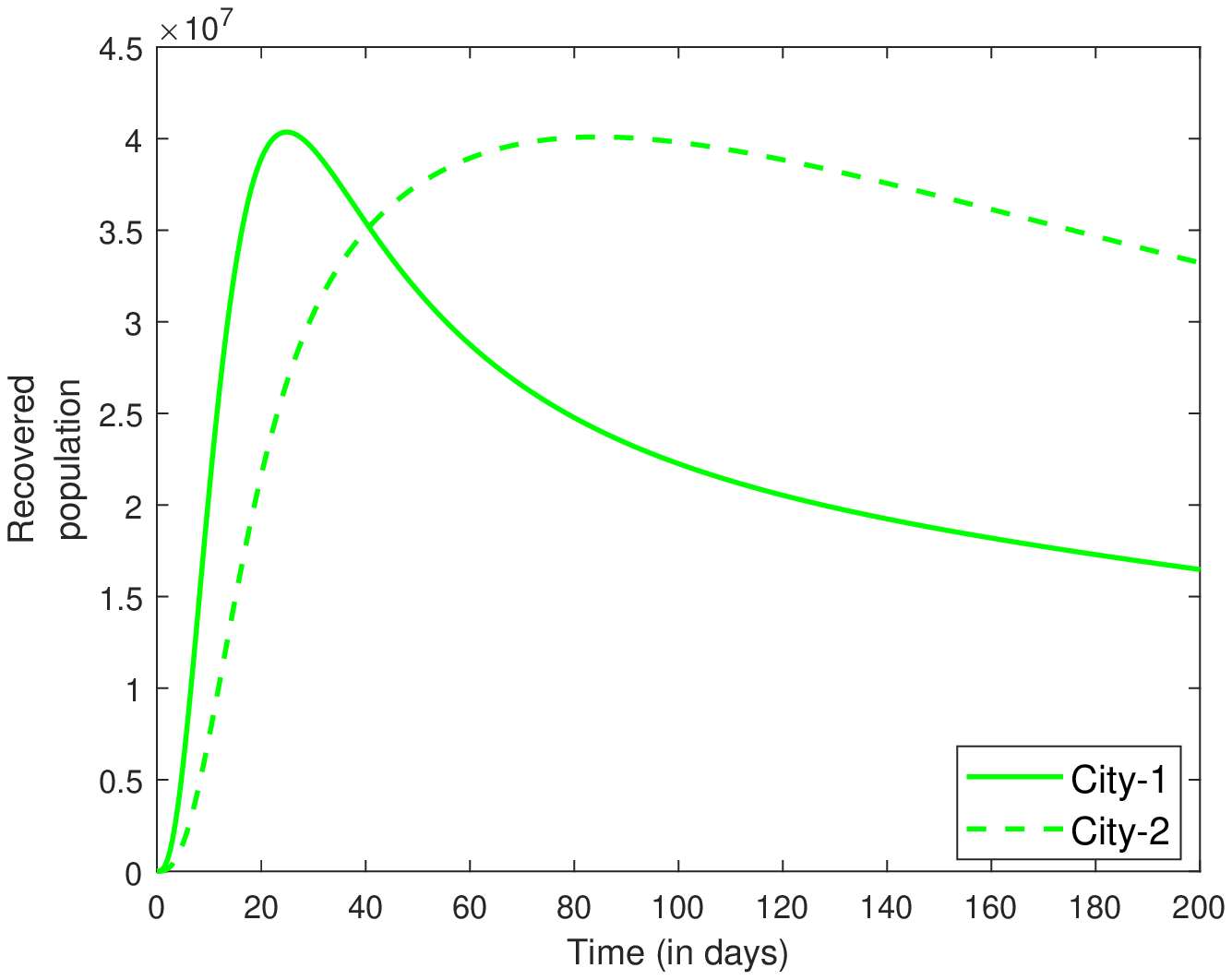}
		\caption{Evolution of recovered population}
		\label{fig:re_mig_1}
	\end{subfigure}
	\caption{Evolution of infected and recovered population for migration rate $m_{12}= 0.01$ and $m_{21}=0.02$}
	\label{fig:mig_1}
\end{figure}

\begin{figure}[!h]
	\centering
	\begin{subfigure}{.5\textwidth}
		\centering
		\includegraphics[scale=0.55]{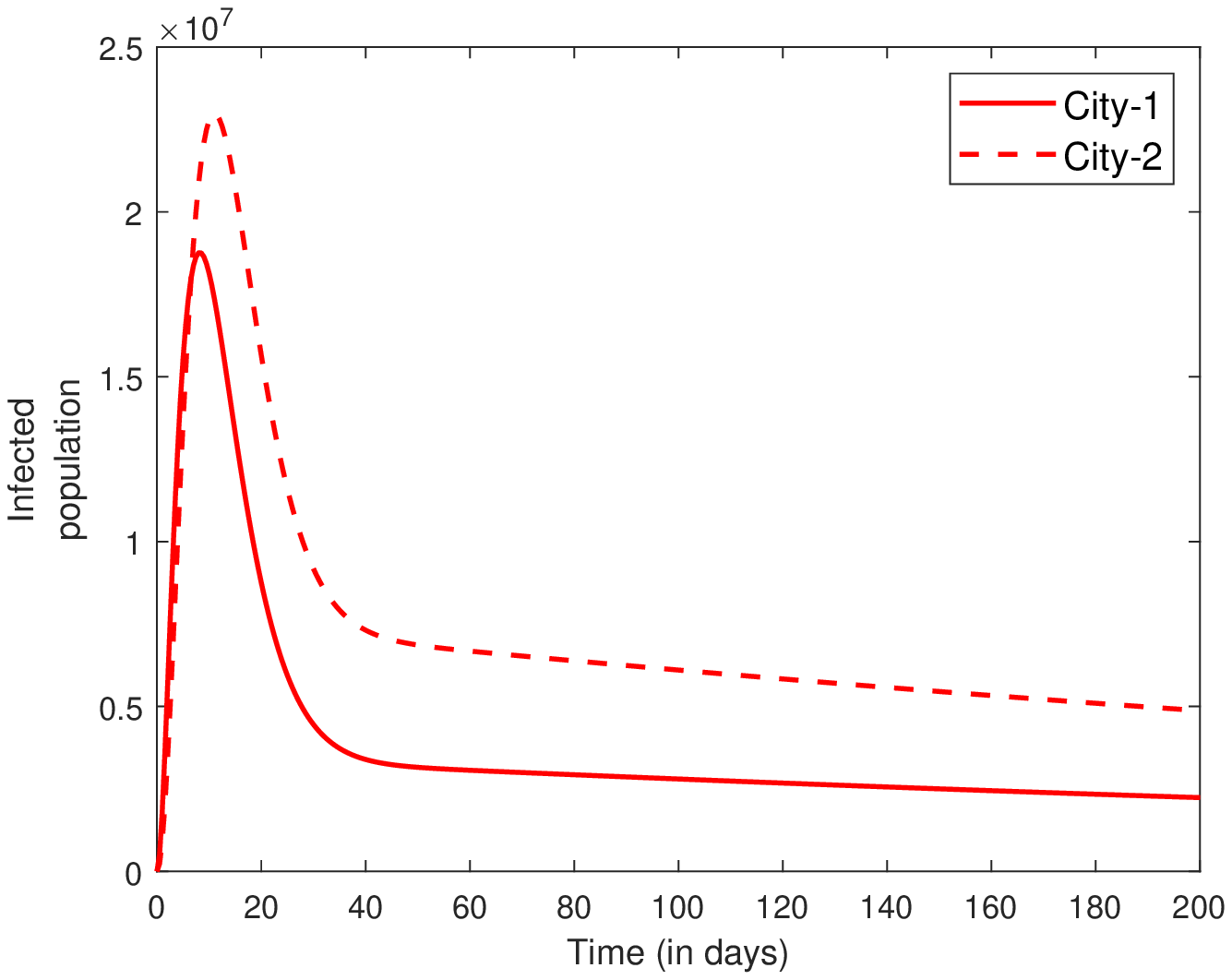}
		\caption{evolution of infected population. The maximum infected population for city 1 and city 2: $(I^{m}_{1}, I^{m}_{2}) = (1.92 \times 10^7 , 2.27 \times 10^7)$}
		\label{fig:inf_mig_2}
	\end{subfigure}%
	\begin{subfigure}{.5\textwidth}
		\centering
		\includegraphics[scale=0.55]{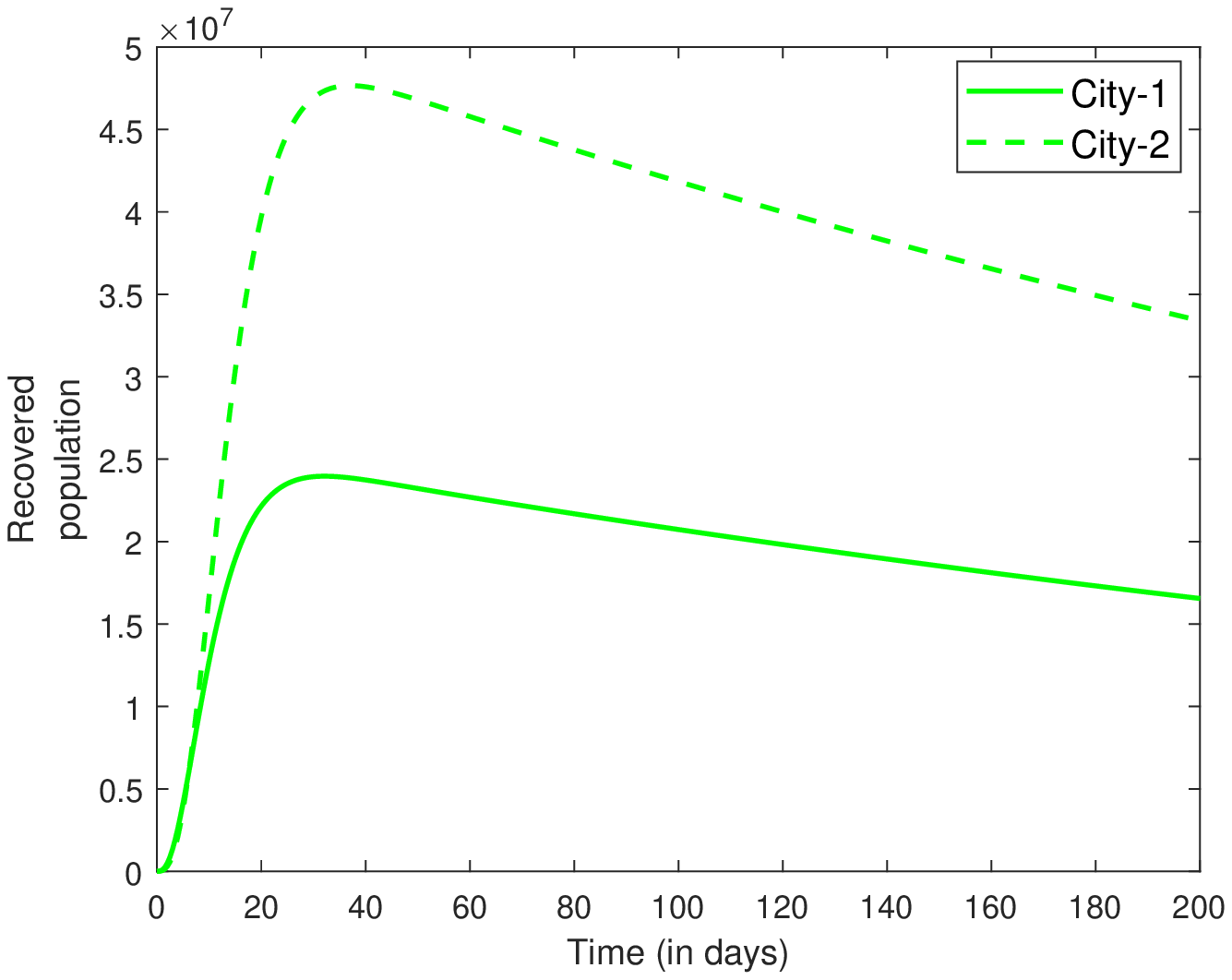}
		\caption{Evolution of recovered population}
		\label{fig:rec_mig_2}
	\end{subfigure}
	\caption{Evolution of infected and recovered population for migration rate $m_{12}= 0.05$ and $m_{21}=0.1$}
	\label{fig:mig_2}
\end{figure}

\begin{figure}[h]
	\centering
	\begin{subfigure}{.5\textwidth}
		\centering
		\includegraphics[scale=0.55]{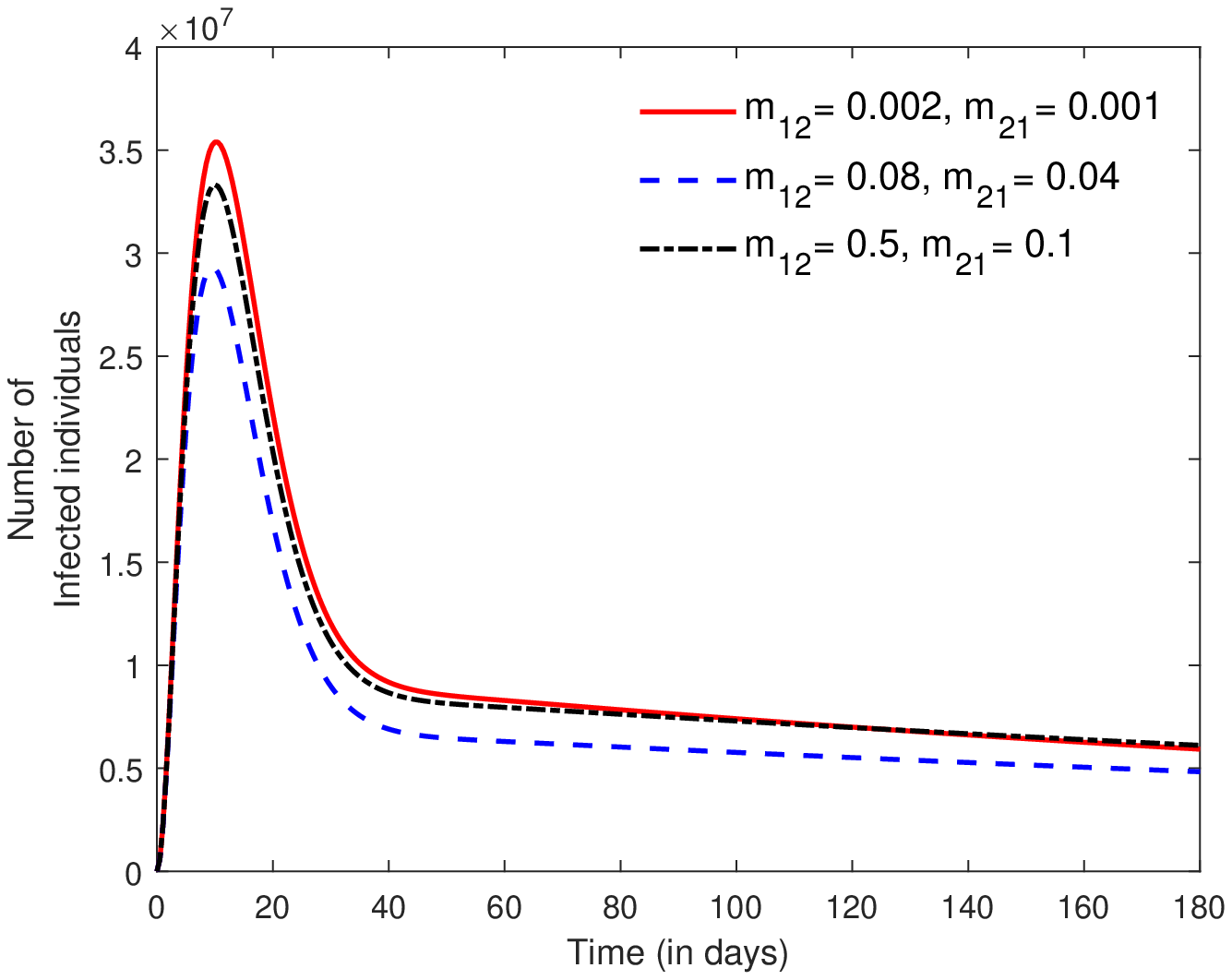}
		\caption{Evolution of infected population for city 1}
		\label{fig:inf_mig_city1}
	\end{subfigure}%
	\begin{subfigure}{.5\textwidth}
		\centering
		\includegraphics[scale=0.55]{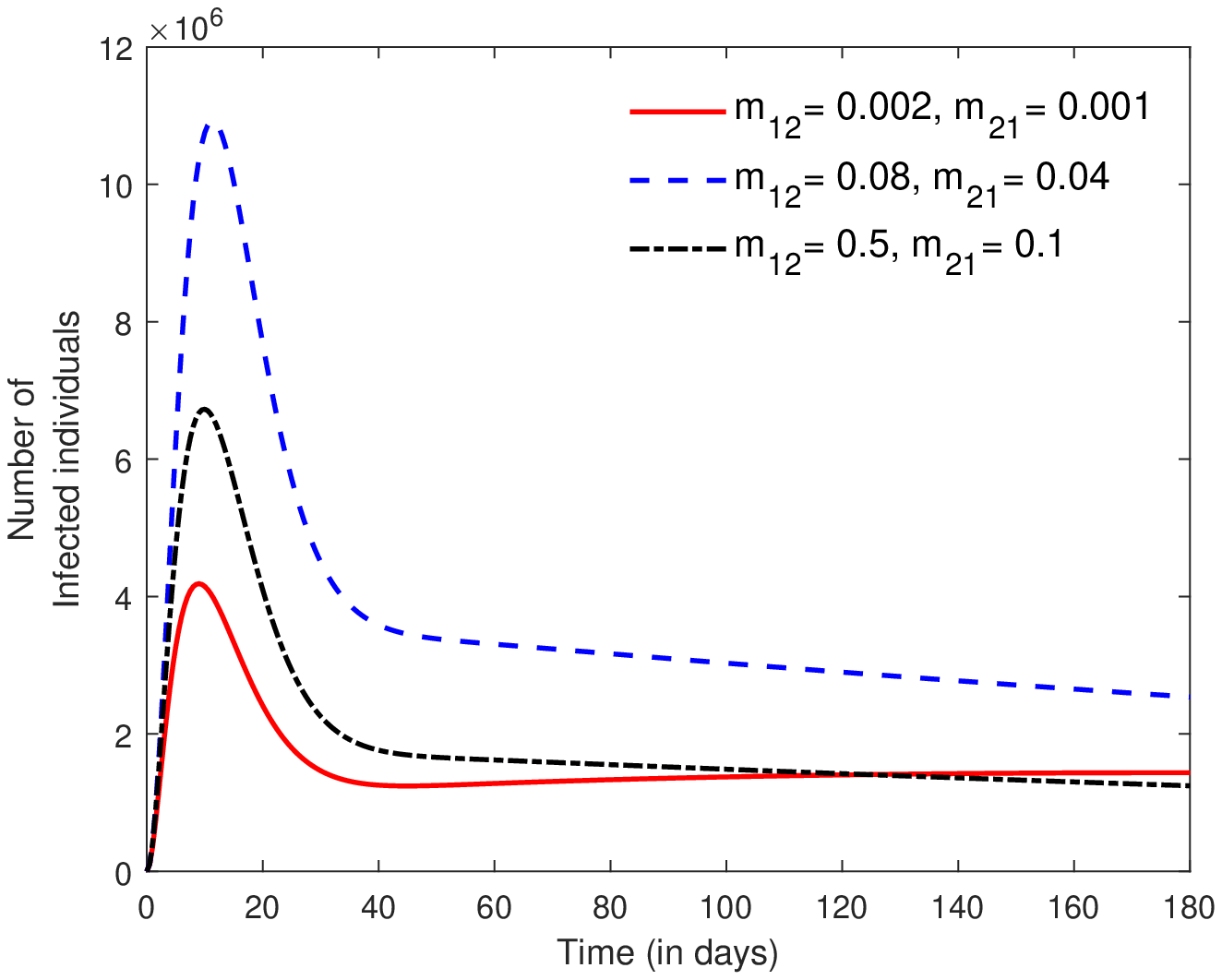}
		\caption{Evolution of infected population for city 2}
		\label{fig:inf_mig_city2}
	\end{subfigure}
	\caption{Evolution of infected population for both cities for various migration rates.}
	\label{fig:mig_var}
\end{figure}

\begin{figure}[!h]
	\centering
	\begin{subfigure}{.5\textwidth}
		\centering
		\includegraphics[scale=0.55]{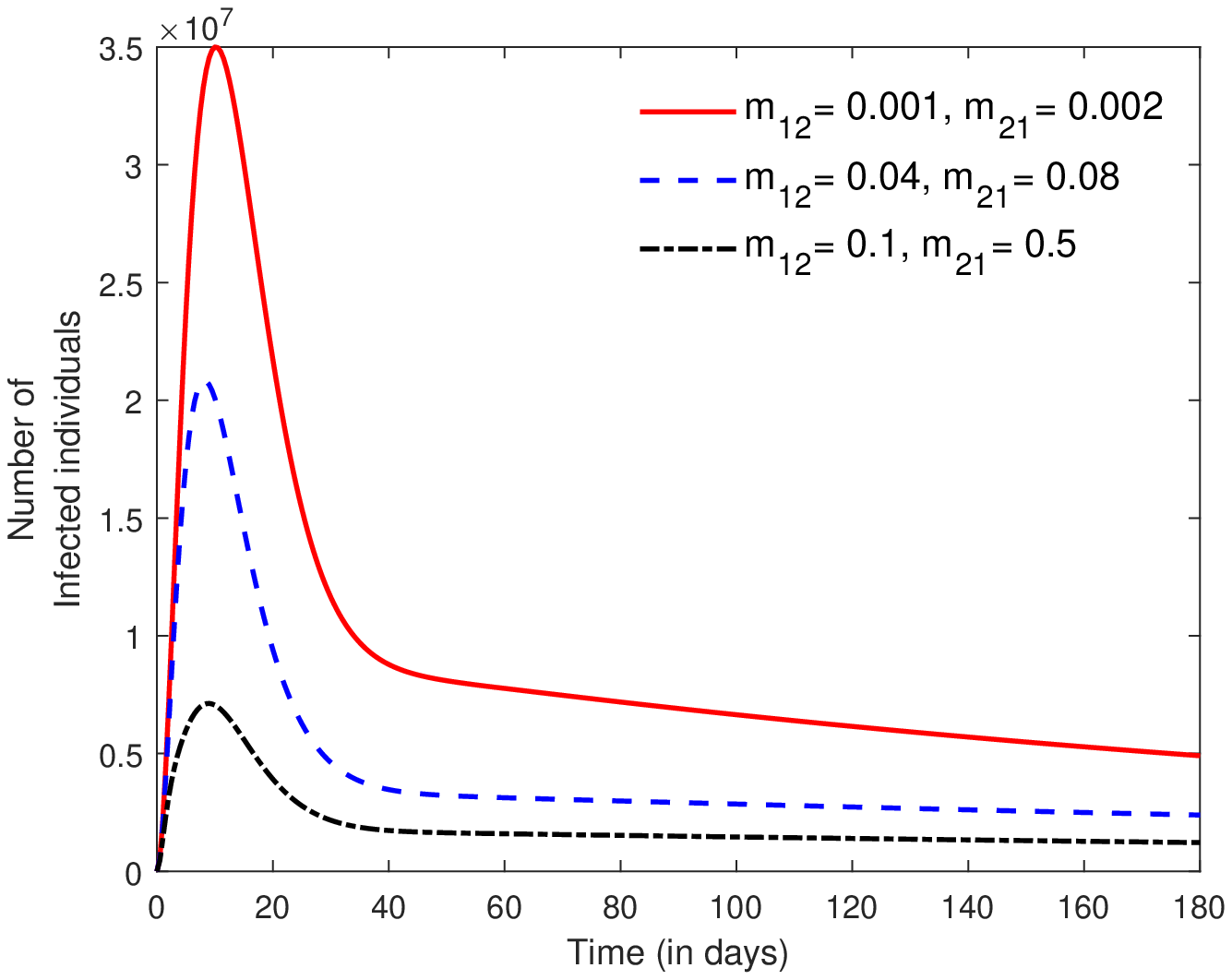}
		\caption{Evolution of infected population for city 1}
		\label{fig:rev_inf_mig_city1}
	\end{subfigure}%
	\begin{subfigure}{.5\textwidth}
		\centering
		\includegraphics[scale=0.55]{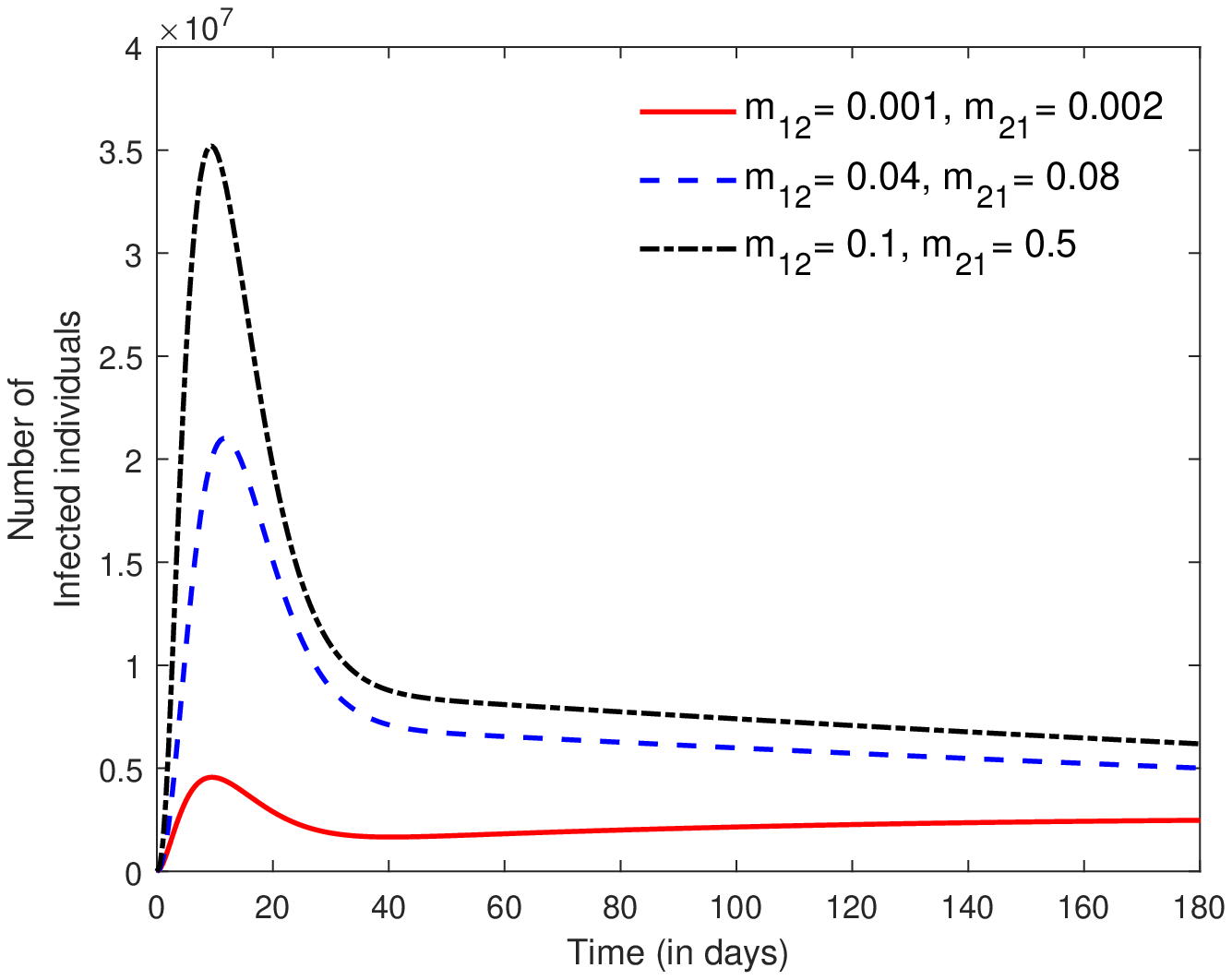}
		\caption{Evolution of infected population for city 2}
		\label{fig:rev_inf_mig_city2}
	\end{subfigure}
	\caption{Evolution of infected population for both cities for various (reverse)migration rates.}
	\label{fig:rev_mig_var}
\end{figure}

In this section, we present some numerical simulations regarding the application of population migration between two cities. As explained earlier, city-1 is the larger city(also denoted by L) with a population of $10^8$, and city-2 is the smaller city (also denoted by S) has a population of $10^7$. We allow greater migration from city L to city S. The parameters used for simulation is summarized in Table \ref{Tabel_2} and the inital conditions used for cities is mentioned in sec.\ref{sec:description_of_cities}.

Starting with the thousand exposed and one infected individual for both the cities, we analyze the effect of migration on both the cities. Figure(\ref{fig:mig_1} )shows the effect of migration from city L to city S. We observe as shown in fig.(\ref{fig:inf_mig_1}) that at around day-40 point the infection load on city S becomes greater than the city L and infected population increases and almost saturates at 4.77 $\times 10^6$ infected individuals. As the number of infected population decreases for city L the number of recovered individuals comes down too, as shown in fig.(\ref{fig:re_mig_1}). We observe that even a small amount of migration ($m_{21} = 0.02$) to a smaller city can eventually equalize and even increase the infected population in the city. Further, if we allow a larger migration from city L to city S as shown in fig. (\ref{fig:mig_2}), infection load in the smaller city takes over the larger city, and a maximum number of the infected population is 2.27 $\times 10 ^7$ as compared 1.92 $\times 10 ^7$ of city L (see fig.(\ref{fig:inf_mig_2})).

Effect of different migration rates is shown in fig.(\ref{fig:mig_var}) and (\ref{fig:rev_mig_var}). We analyze the effect of migration (higher rates from city-2 to city-1) and reverse migration (higher rates from city-1 to city-2). For the case of migration, we see that city-1, which accepts the migrants from city-2 at a higher rate, has a higher infectious individual. For migration rate $m_{12}= 0.002$ and  $m_{21}= 0.001$, the maximum infected individual in city-1 and city-2 are $3.54 \times 10^7 $ and $4.19 \times 10^6 $ respectively. As we increase the migration rate to $m_{12}= 0.08$ and  $m_{21}= 0.04$, we see a decrease in maximum infected individual in city-1 and increase in city-2 which is $2.93 \times 10^7 $ and $1.09 \times 10^7 $ respectively. This is because now more individuals enter into city-2, thus raising the number of infected individuals.

For the case of reverse migration we observe that for smaller migration rate $m_{12}= 0.001$ and  $m_{21}= 0.002$ the infection load is still higher in the city which migrants leave (city-1) to the other city (city-2). As we increase the migration rate the number of infected individual in city-2 increases. For instance as shown in fig.(\ref{fig:rev_inf_mig_city2}), for $m_{12}= 0.1$ and  $m_{21}= 0.5$, the maximum number of infected individual in city-1 and city-2 is $7.13 \times 10^6 $ and $3.52 \times 10^7 $ respectively.

\newpage
\subsection{Punctuated Lockdown}

\begin{figure}[!h] 
	\centering
	\includegraphics[scale=1.]{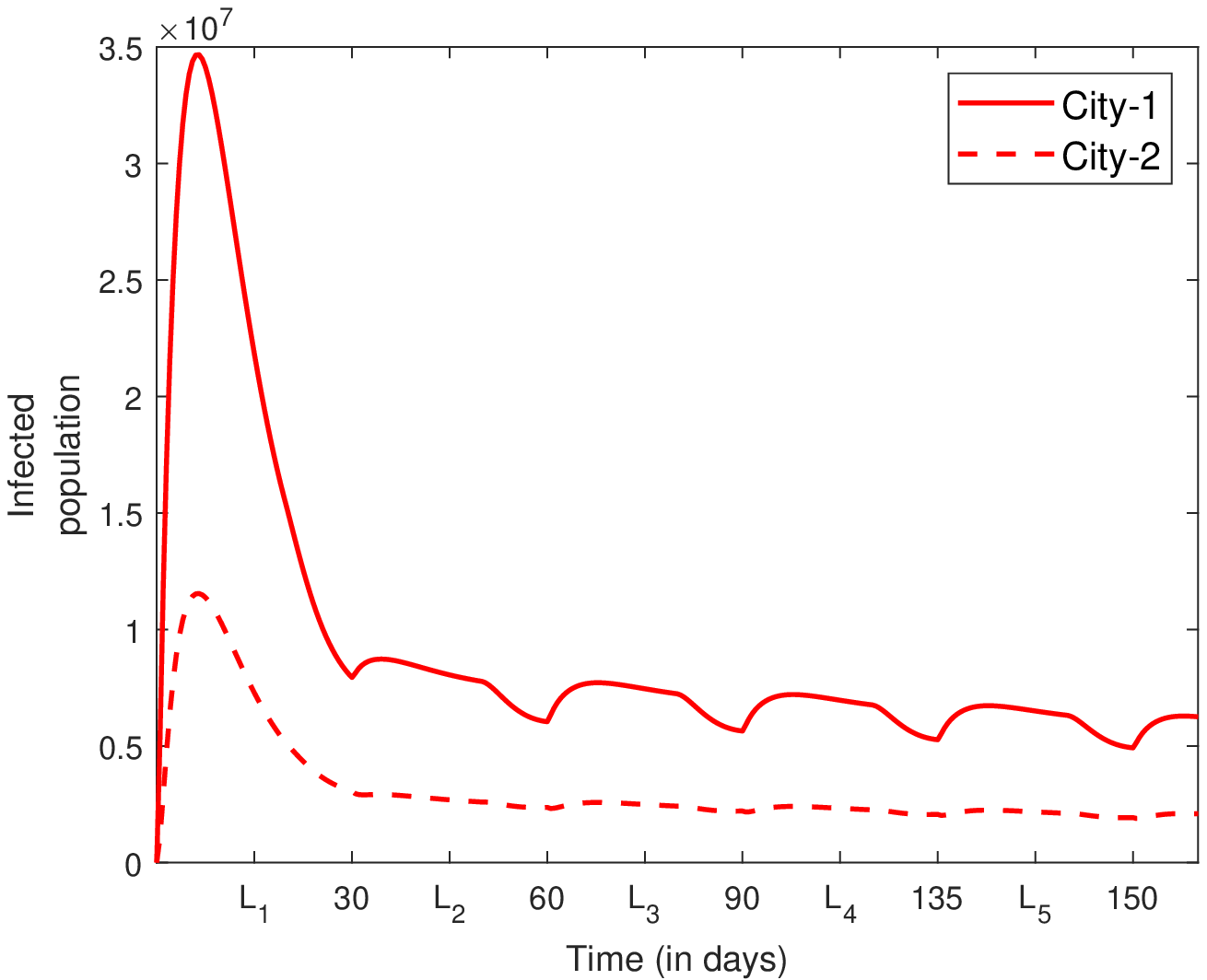}
	\caption{Evolution of infected population under punctuated lockdown. }
	\label{fig:punc_lock}
\end{figure}%

In this section, we study the effect of punctuated lockdown in both cities. Instead of relaxing the lockdown completely, the restrictions are lifted for 20 days and then imposed back for the next ten days. For the relaxed phase  $(\eta_1, \eta_2, m_1, m_2) = (0.05, 0.025,0.75,0.25)$ and for lockdown phase $(\eta_1, \eta_2, m_1, m_2) = (0.9, 0.5,0.05,0.02)$ are used. A damped oscillatory behavior is observed if a small fraction does turn positive again under Covid testing after recovery from the infection. The infected populations reach the peak for both the cities and then decrease to rise to certain values before decreasing again. The infected population decreases very gradually but does not come down to zero. This can be due to the inflow of population per unit time given by A (see fig.(\ref{fig:punc_lock})).

The oscillatory behavior in the population is the result of a small fraction of recovered individuals becoming susceptible again. This implies that one should keep a close observation over even such individuals who are recovered. The importance of punctuated lockdown is highlighted because the dips during the lockdown period allow the health infrastructure to prepare for the ensuing surge in the infection.

\section{Population Migration: Application to three cities} \label{sec:three_city}

In this section we apply the system(\ref{eqn: pop_model}) to analyse the migration among three cities. In the following section, we see the variation of reproduction number $R_0$ for the following infectious sub-system.

\begin{align}
	\label{eqn: sub_system_three}
	\frac{dE_{1}}{dt} &=\frac{\beta S_{1}I_{1}}{1+ \eta_{1} I_{1}} -\sigma E_{1} -d E_{1} -m_{21}E_{1} - m_{31}E_{1}  +m_{13}E_{3} + m_{12}E_{2} \nonumber \\
	\frac{dI_{1}}{dt} &= \sigma E_{1} -\gamma I_{1} -\alpha I_{1}-dI_{1}-m_{21}I_{1} - m_{31}I_{1}  +m_{13}I_{3} + m_{12}I_{2}\nonumber  \\ \nonumber \\
	\frac{dE_{2}}{dt} &=\frac{\beta S_{2}I_{2}}{1+ \eta_{2} I_{2}} -\sigma E_{2} -d E_{2} -m_{32}E_{2} - m_{12}E_{2}+m_{21}E_{1} + m_{23}E_{3}  \\
	\frac{dI_{2}}{dt} &= \sigma E_{2} -\gamma I_{2} -\alpha I_{2}-dI_{2}-m_{32}I_{2} - m_{12}I_{2}+m_{21}I_{1} + m_{23}I_{3} \nonumber \\ \nonumber \\
	\frac{dE_{3}}{dt} &=\frac{\beta S_{3}I_{3}}{1+ \eta_{3} I_{3}} -\sigma E_{3} -d E_{3} -m_{13}E_{3} - m_{23}E_{3} +m_{31}E_{1} + m_{32}E_{2} \nonumber \\
	\frac{dI_{3}}{dt} &= \sigma E_{3} -\gamma I_{3} -\alpha I_{3}-dI_{3}-m_{13}I_{3} - m_{23}I_{3} +m_{31}I_{1} + m_{32}I_{2}  \nonumber
\end{align}

\subsection{Reproduction Number $R_0$}

We outline the procedure to calculate the reproduction number for three cities in this section. The procedure is similar to that used in section (\ref{sec: R_0}) following the method outlined by Diekmann \textit{et.al} \cite{diekmann2010construction} and Arino \textit{et.al} \cite{arino2015epidemiological} to find the reproduction number ($R_0$) by constructing next-generation matrix(NGM) for the complete system.

For the computation of $R_0$, we only regard the states that apply to infected individuals, which for the case of three cities are $(E_1, E_2, E_3, I_1, I_2, I_3)$. For notations used see Table \ref{Table_1}. Unlike the case of two cities, expressing the reproduction number in analytical form is difficult for this case, and hence reproduction number is calculated numerically using MATLAB 2020(a) \cite{MATLAB:2020}.

The matrix \textbf{T}, describing the production of new infections and \textbf{$\Sigma$}, describing the changes in the state is:
$$
\mathbf{T} =
\begin{pmatrix}
	0 & \nicefrac{\beta A}{m_{21}+m_{31}+d} & 0 & 0& 0 & 0 \\
	0 & 0 & 0 & 0& 0 & 0\\
	0 & 0 & 0 & \nicefrac{\beta A}{m_{12}+m_{32}+d}& 0 & 0 \\
	0 & 0 & 0 & 0& 0 & 0\\
	0 & 0 & 0 & 0& 0 &\nicefrac{\beta A}{m_{13}+m_{23}+d}\\
	0 & 0 & 0 & 0& 0 & 0
\end{pmatrix}
$$

\begin{figure}[!h] 
	\centering
	\includegraphics[scale=1]{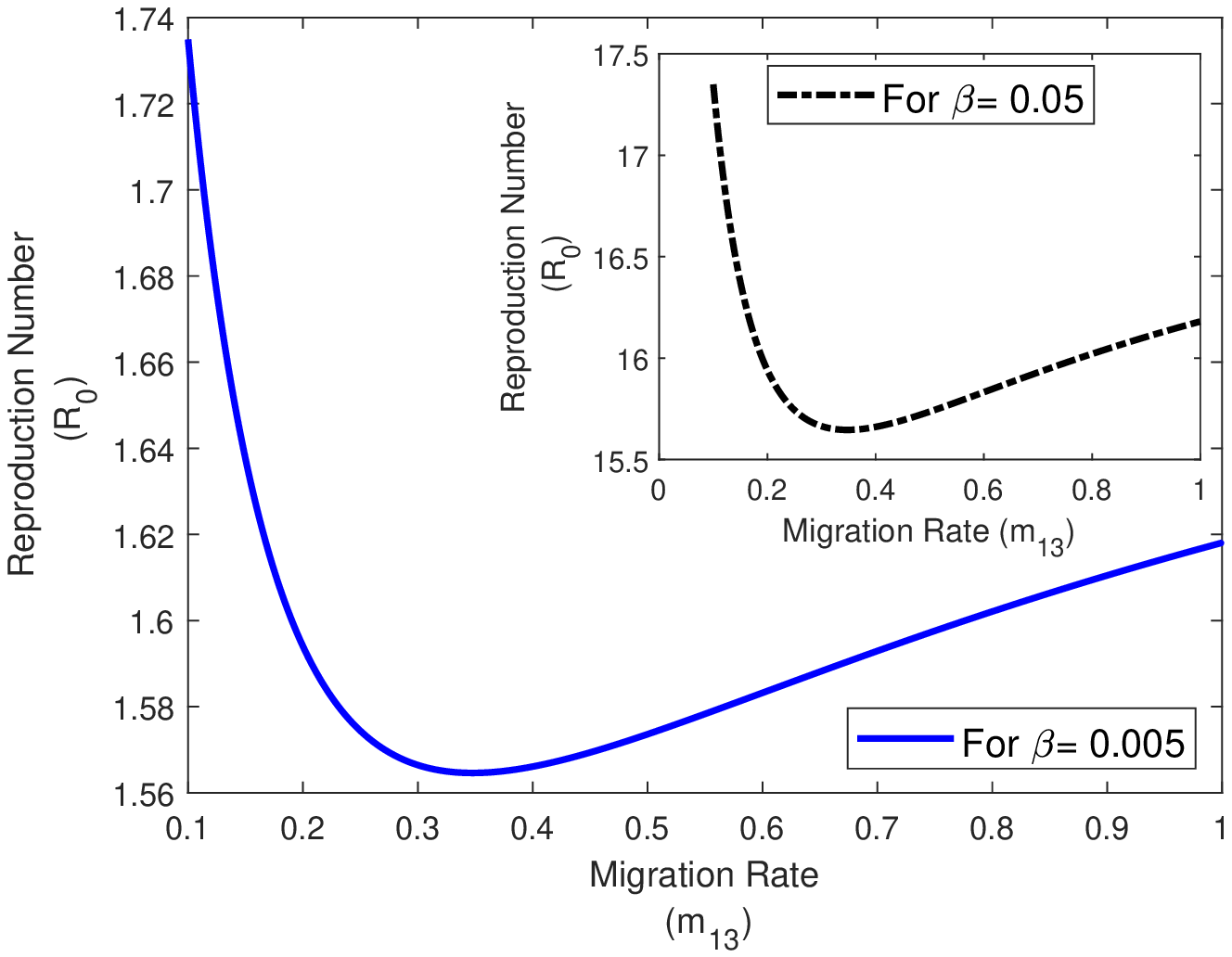}
	\caption{Variation of $R_0$ with migration parameter $m_{13}$.The usual parameters used are the same as in Table \ref{Tabel_2} and migration parameters used are $m_{12}$ = $m_{21}$ = $m_{31}$ = $m_{23}$ = $m_{32}$ = $0.1$}
	\label{fig:R0_m13_inset}
\end{figure}%

and,

$$\Sigma = \mbox{\large$
	\begin{pmatrix}
		\Sigma_{11} & 0 & m_{12} & 0  & m_{13} & 0\\
		\sigma & \Sigma_{22} & 0 & m_{12}  & 0 & m_{13} \\
		m_{21} & 0 &  \Sigma_{33} & 0  & m_{23} & 0 \\
		0 & m_{21} & \sigma & \Sigma_{44} & 0 & m_{23}\\
		m_{31} & 0 & m_{32}  & 0 & \Sigma_{55}    & 0  \\
		0 & m_{31} & 0  & m_{32}& \sigma  & \Sigma_{66}
	\end{pmatrix}$}$$

Where,

\begin{align*}
	\Sigma_{11} =& -(\sigma+d+m_{21}+m_{31})\\
	\Sigma_{22} =& -(\gamma +\alpha +d+m_{21}+m_{31})\\
	\Sigma_{33} =& -(\sigma+d+m_{12}+ m_{32} )\\
	\Sigma_{44} =& -(\gamma +\alpha +d+m_{12}+ m_{32} ) \\
	\Sigma_{55} =& -(\sigma+d+m_{13}+ m_{23}) \\
	\Sigma_{66} =& -(\gamma +\alpha +d+m_{13}+m_{23})  \\
\end{align*}

Reproduction number is then defined as $ R_0 = \rho(-T\Sigma^{-1}),$ where $\rho$ denotes the spectral radius, which is the dominant eigenvalue of the matrix $-T\Sigma^{-1}$. The variation of $R_0$ with migration rates is shown in fig.(\ref{fig:R0_m13_inset}) for $\beta=0.005$. It is observed that as the number of cities increases, the effective contact rate ($\beta$) has to be controlled for keeping the reproduction number under check. The figure in inset shows the variation of $R_0$ with migration rate $m_{13}$ for $\beta=0.05$ and $R_0$ is unacceptably large there. The value of $R_0$ at first decreases with an increase in migration rate in general, only to increase rapidly for a higher migration rate. A similar trend is observed for other migration parameters. If we increase the contact rate, the reproduction number increases as well. This highlights the importance of lockdown to curb the infection spread in the cities.

\begin{figure}[!h] 
	\centering
	\begin{subfigure}{.5\textwidth}
		\centering
		\includegraphics[scale=0.55]{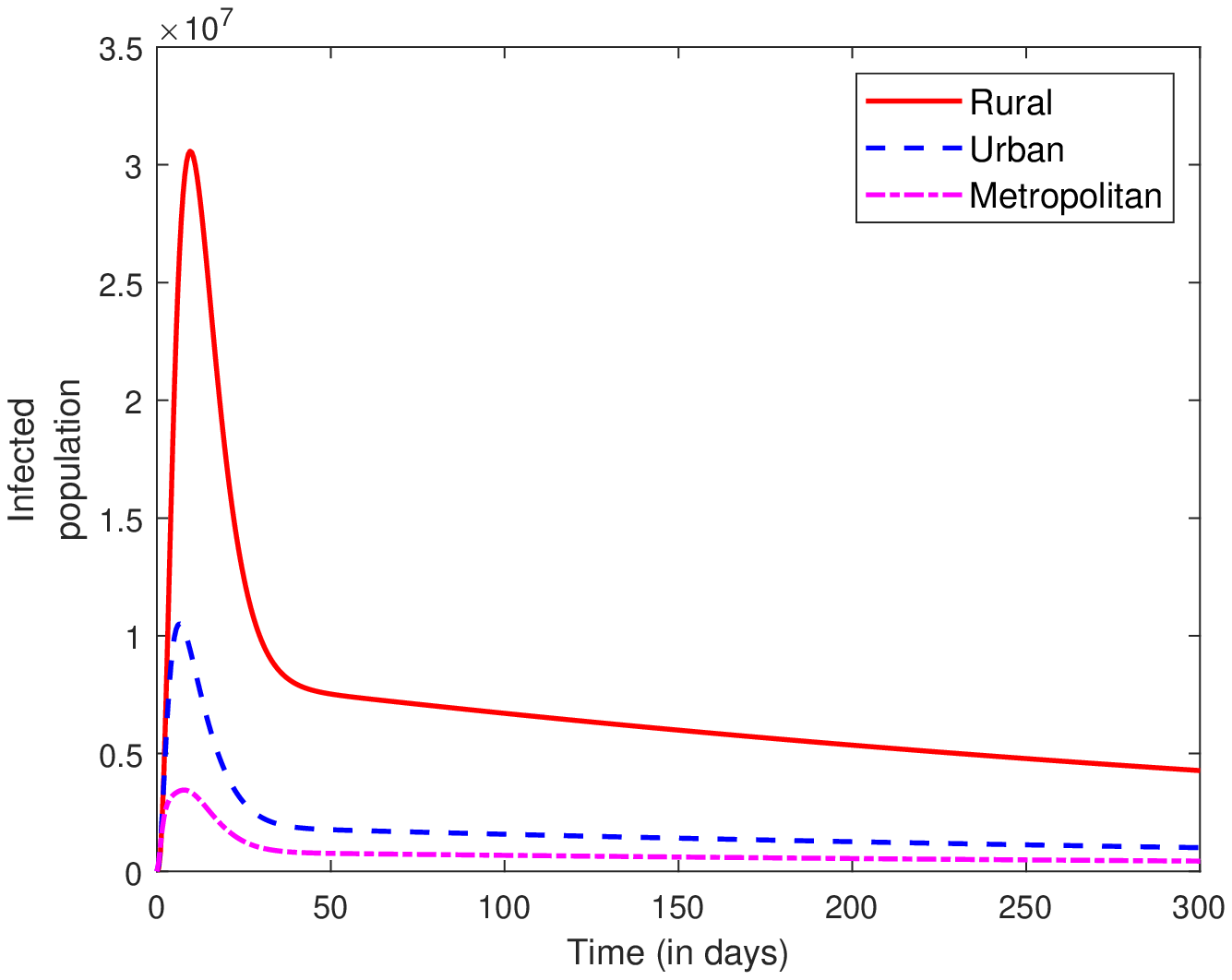}
		\caption{evolution of infected population. The maximum infected population for rural, urban and metropolitan city: $(I^{m}_{R}, I^{m}_{U},I^{m}_{M}) = (3.05 \times 10^7 , 1.05 \times 10^7,  3.45 \times 10^6)$}
		\label{fig:mig3City}
	\end{subfigure}%
	\begin{subfigure}{.5\textwidth}
		\centering
		\includegraphics[scale=0.55]{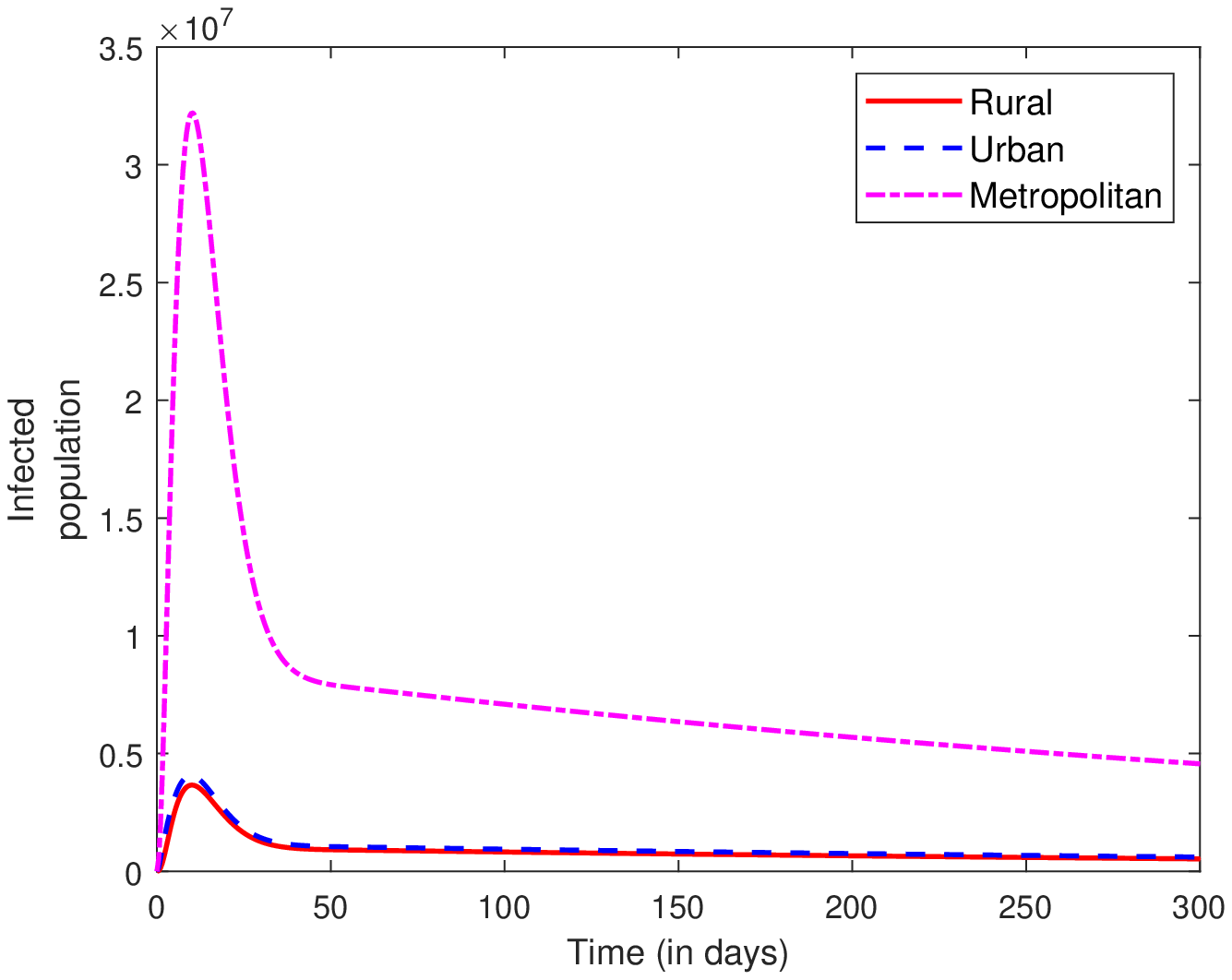}
		\caption{evolution of infected population. The maximum infected population for rural, urban and metropolitan city: $(I^{m}_{R}, I^{m}_{U},I^{m}_{M}) = (3.66 \times 10^6 , 4.06 \times 10^6,  3.22 \times 10^7)$ }
		\label{fig:rev_mig}
	\end{subfigure}
	\caption{Evolution of infected population }
	\label{fig:mig_3_cities}
\end{figure}

\subsection{Effect of migration among cities}

\begin{table}[h]
	\caption{Parameters used for generating fig. \ref{fig:mig_3_cities}}
	\begin{minipage}{.5\textwidth}
		\begin{tabular}{cccc}
			\hline
			parameter &Value & parameter &Value \\
			\hline
			$m_{12}$ & 0.15 & $m_{21}$ & 0.02 \\
			$m_{13}$ & 0.35 &$m_{31}$ & 0.05 \\
			$m_{23}$ & 0.25 & $m_{32}$ & 0.05 \\
			\hline
		\end{tabular}
	\end{minipage}%
	\begin{minipage}{0.5\textwidth}
		\begin{tabular}{cccc}
			\hline
			parameter &Value & parameter &Value\\
			\hline
			$m_{12}$ & 0.05 & $m_{21}$ & 0.15 \\
			$m_{13}$ & 0.05 & $m_{31}$ & 0.35 \\
			$m_{23}$ & 0.02 & $m_{32}$ & 0.25 \\
			\hline
		\end{tabular}
	\end{minipage}
	\label{Table_3}
\end{table}

As an application to the system (\ref{eqn: pop_model}), in this section, we present the numerical simulation for three cities. We divide the cities into three types to study the effect of migration on systems with a different demography. The first type is a smaller (call it rural) city with a small population($N_0= 10^6$) and individuals who primarily work in more developed cities. We assume that the city has few preventive measures. The second type is a bigger city (say, urban) with a population ($N_0=10^7$) with better facilities than the rural one. Individuals in the urban city too move to even larger and more developed cities for better work perspectives. We assume urban city follows preventive measures up to a certain extent. Finally, the third type is a metropolitan city with a large population ($N_0= 10^8$) with greater facilities and better implementation of preventive measures.

First, we analyze the effect of migration (higher migration rates from a larger city to a smaller city) from a metropolitan city to an urban and rural city (see fig. \ref{fig:mig_3_cities}). The parameters used to generate fig. \ref{fig:mig_3_cities} are shown in Table \ref{Table_3}. We observe from fig. \ref{fig:mig3City} that starting from the same number of infected individuals, the number of infected population for the rural city increases rapidly, reaching a maximum of $3.06 \times 10^7$ as compared to $0.34 \times 10^7$ infected individuals in the metropolitan city. This puts a tremendous load on the rural and urban cities. Another important thing to observe is that the infected population for the overburdened city does not come down to zero. Next, we allow for the reverse migration to the developed city, as shown in fig. \ref{fig:rev_mig}. We see that even though individuals migrate to the urban as well as a metropolitan cities, the infection overload is in a metropolitan city where the number of infected individuals reaches $3.22 \times 10^7$

\subsection{Punctuated Lockdown}

\begin{figure}[h] 
	\centering
	\includegraphics[scale=1]{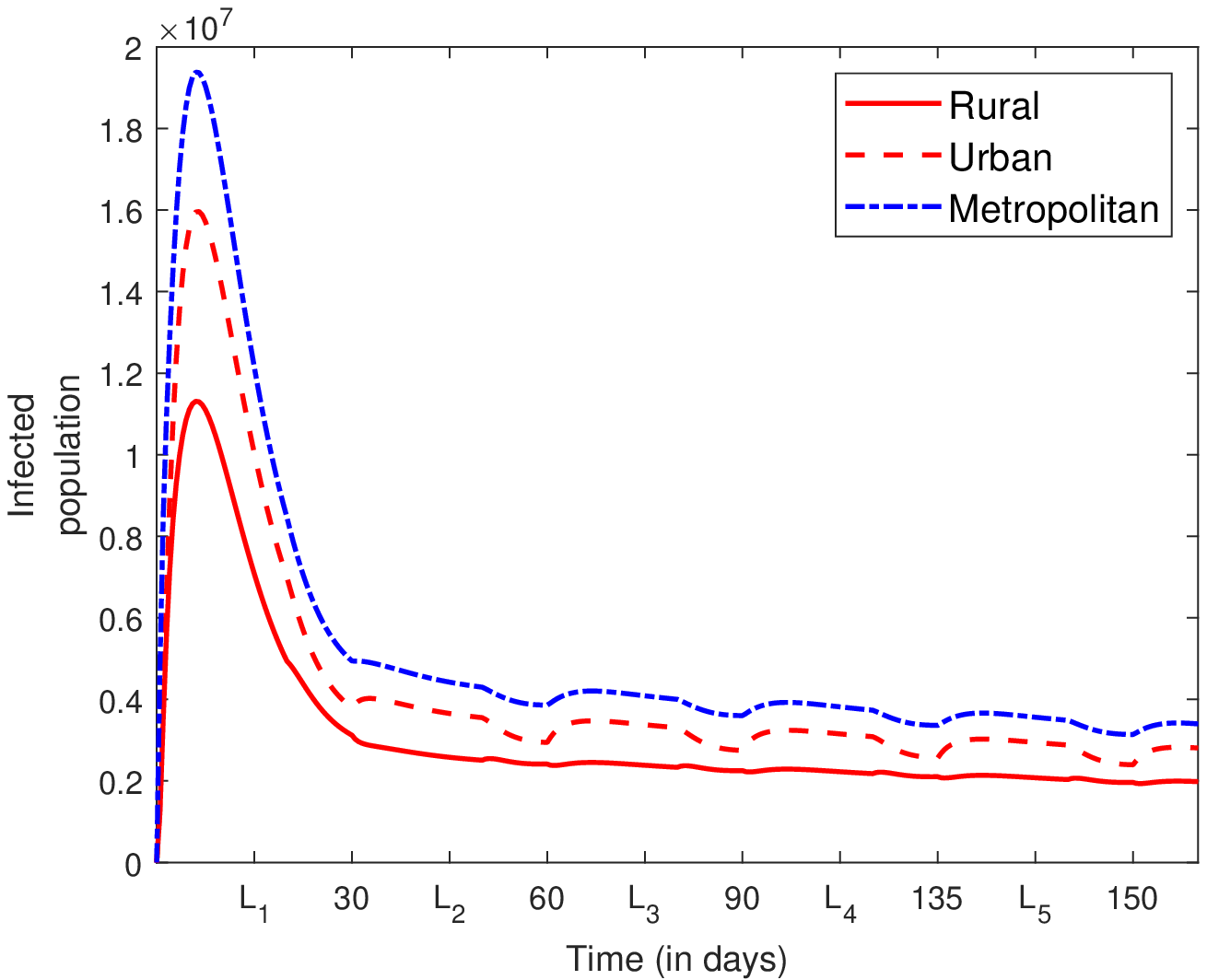}
	\caption{Evolution of infected population under punctuated lockdown.}
	\label{fig:punc_lock_3_city}
\end{figure}%

Similar to the punctuated lockdown for two cities, we study the effect of punctuated lockdown for three cities. Instead of relaxing the lockdown completely, the restrictions are lifted for 20 days and then imposed back for the next ten days. Assuming a small fraction turns positive again after recovering from the infection, an oscillatory behavior is also observed. The infected populations reach the peak for the cities and then decrease to rise to specific values before reducing again. The infected population decreases very gradually but does not come to zero.

The important thing observed here is the effect of $\delta$ on the punctuated lockdown. If the immunity developed is for a lifetime and no recovered individual becomes infected again, then the number of infected individuals comes down to zero. The oscillatory behavior in the population is the result of a small fraction of recovered individuals becoming susceptible again. This leads to the idea that one should have close observation over individuals who are recovered.

The idea of punctuated lockdown is vital in the sense that it gives the necessary time to recover the overstretched health infrastructure for the ensuing increase in infections. Due to the nature of the virus, a large section of the population is bound to get infected. Still, it is crucial to allow the health infrastructure to get back on its feet during the lockdown to take care of the new infections generated after it is relaxed.

\newpage

\section{Conclusion}

We considered the SEIRS epidemic diffusion model with population migration for $n$ cities. Then as an application to the model, we analyzed migration between two and three cities. The migration rate between cities is unequal, and individuals from all compartments can migrate between the cities. The reproduction number has been calculated, and its variation with migration rate is studied. Unless a vaccine is forthcoming in the immediate future, protective health measures are paramount to prevent the spread. In order to control the infected population, preventive measures should be implemented at lower thresholds.

A punctuated lockdown is also studied in this context. The oscillatory behavior is observed when a small fraction of recovered individuals are allowed to become susceptible again. This raises the critical point that the recovered individuals should be observed to prevent the increase of susceptibles. An increase in the number of susceptibles pushes up the fraction of infected individuals leading to consequent peaks. The punctuated lockdown is an exciting deviation from the extant strategies and helps handle situations post-first lockdown. The implementation of punctuated lockdown provides necessary recovery time for the health infrastructure to accommodate the upcoming infection overload.

The model with different strategies emphasizes finding ways to minimize the spread until a cure or a vaccine is available.

\bibliographystyle{unsrt}
\bibliography{References}
\end{document}